# Throughput Enhancement Using Multiple Antennas in OFDM-based Ad Hoc Networks under Transceiver Impairments

Pengkai Zhao, and Babak Daneshrad

Wireless Integrated Systems Research (WISR) Group, Electrical Engineering Department,

University of California, Los Angeles, CA 90095 USA

(e-mail: pengkai@ee.ucla.edu; babak@ee.ucla.edu)

## Abstract

Transceiver impairments, including phase noise, residual frequency offset, and imperfect channel estimation, significantly affect the performance of Multiple-Input Multiple-Output (MIMO) system. However, these impairments are not well addressed when analyzing the throughput performance of MIMO Ad Hoc networks. In this paper, we present an analytical framework to evaluate the throughput of MIMO OFDM system under the impairments of phase noise, residual frequency offset, and imperfect channel estimation. Using this framework, we evaluate the Maximum Sum Throughput (MST) in Ad Hoc networks by optimizing the power and modulation schemes of each user. Simulations are conducted to demonstrate not only the improvement in the MST from using multiple antennas, but also the loss in the MST due to the transceiver impairments. The proposed analytical framework is further applied for the distributed implementation of MST in Ad Hoc networks, where the loss caused by impairments is also evaluated.

# I. Introduction

Wireless Ad Hoc networks are considered as an important part of next generation wireless communication systems. Since there is no centric controller in the network, the performance is seriously constrained by the co-channel interference from other users. Multiple antennas, which can mitigate the interference and increase the throughput, constitute a prospective way for improving the throughput performance of Ad Hoc networks. They have already been proposed and discussed in several works [1-5].

In this paper, we focus on the throughput enhancement from the usage of multiple antennas in Ad Hoc networks. In this area, the authors in [1-3, 7] investigate the sum throughput in a MIMO enabled Ad Hoc networks by discussing and comparing different MIMO techniques, including spatial multiplexing, beam-forming, etc. Meanwhile, the achievable throughput bound under MIMO system is evaluated in [4] via the linear programming method, where the topology information among the nodes are utilized. Moreover, considering the adaptive management of the transmit power, distributed schemes using game theory is proposed in [5, 6] for further increasing the throughput of MIMO Ad Hoc network. Finally, field test results of MIMO systems in realistic Ad Hoc networks are provided in [8] and compared to the Shannon Capacity. In these works, the throughput performance of MIMO system is mainly scaled from the Shannon Capacity. However, due to some practical factors in the physical layer, especially the impairments in the transceiver, the performance of realistic MIMO systems is always far away from the Shannon Capacity. Thus it is necessary to take the transceiver impairments into account when analyzing the network performance.

This paper attempts to provide an analytical framework to evaluate the impacts of transceiver impairments on the throughput enhancement of MIMO Ad Hoc networks. The OFDM modulation scheme is assumed at every node, which provides an efficient method for utilizing the wide-band frequency resource. The discussed impairments are both from RF domain and from Base-band domain. With respect to the impairments in RF domain, nonlinearity in power amplifier, phase noise and I/Q imbalance have been conventionally discussed [9]. Previous works have shown that the nonlinearity problem can be solved by the back-off and pre-distortion mechanism [9], [10] (the impact is that the maximum transmit power is constrained), and I/Q imbalance problem can be mitigated by signal processing techniques [11]. However, phase noise is still difficult to be eliminated and is often treated as white noise [12]. Thus we only consider the phase noise impairment in RF domain. With respect to the impairments in base-band domain, we discuss the residual frequency offset that causes ICI interference in the sub-carriers [13], as well as the imperfect channel estimation that affects the performance of MIMO detector [14].

As a result, in this paper, three different transceiver impairments, including phase noise, carrier frequency offset, and imperfect channel estimation, are discussed in a MIMO-OFDM based Ad Hoc networks. We first provide an analytical framework to scale the performance of the MIMO OFDM system under these impairments. Based on this framework, we evaluate the Maximum Sum Throughput (MST) in the network by optimizing the power and modulation method of every user. We use Monte Carlo simulations to evaluate both the increment of MST by multiple antennas, and the loss in MST because of the system impairments.

Furthermore, since distributed control method is naturally required in Ad Hoc networks, we also discuss how the proposed framework can be applied for the distributed implementation of MST. Our proposed framework can provide a series of SINR thresholds for different data rates, thus we provide a two-stage distributed power control method to utilize these results. We finally evaluate the impacts of the transceiver impairments under this distributed power control framework.

The rest of the paper is organized as follows. System model is first introduced in

Section II. RF impairments are discussed in Section III, and Base-band impairments are in Section IV. We provide the proposed analytical framework, as well as the optimization of the sum throughput in the network in Section V. Simulation results are given in Section VI, and we conclude in Section VII.

## II System Model

We consider a one-hop Ad Hoc network with independent transceiver pairs. Every pair is composed of one transmit node and one receive node, sharing the same frequency band. The transmit and receive nodes in the $i$ th transceiver pair are labeled as Tx Node $i$ and Rx Node $i$, respectively. Each pair in the network use the OFDM modulation and $N$ receive antennas, while $M_i$ transmit antennas are employed at Tx Node $i$. The transmitted vector at the $k$ th sub-carrier of Tx Node $i$ is $\mathbf{X}_i(k)$, which is an $M_i \times 1$ vector with zero mean and unit variance in each element.

The wireless channel between Tx Node $i$ and Rx Node $j$ are described by path loss and Rayleigh fading. The path loss follows the simplified model in [15], which is:

$$L_P(d_{ji})(\text{dB}) = L_P(d_0) + 10\alpha \log_{10} \frac{d_{ji}}{d_0} \qquad (1)$$

Here $d_{ji}$ is the distance between Tx Node $i$ and Rx Node $j$, and $d_0$ is the reference distance. We use $\rho_{ji}$ to denote $L_P(d_{ji})$ in decimal, $\rho_{ji} = 10^{-\frac{L_P(d_{ji})}{10}}$. Meanwhile, the Rayleigh fading channel between Tx Node $i$ and Rx Node $j$ on the $k$ th subcarrier is $\mathbf{H}_{ji}(k)$. It is a $N \times M_i$ matrix with independent complex Gaussian variables (zero mean and unit variance for each element). We assume that the transmit power of Tx Node $i$ is $P_i$. It is uniformly allocated among $M_i$ transmit antennas ($P_i/M_i$ for each transmit antenna). $P_i$ is adaptively adjusted subject to a maximum transmit power $P_{\max}$.

Consider that there are $K$ transceiver pairs transmitting simultaneously in the network. Without impairments in the physical layer, the received signal at the $k$ th sub-carrier of Rx Node $j$ is denoted as:

$$\mathbf{Y}_j(k) = \sqrt{\frac{P_j \rho_{jj}}{M_j N_S}} \mathbf{H}_{jj}(k) \mathbf{X}_j(k) + \mathbf{R}_j(k) + \mathbf{N}_j(k) \tag{2}$$

$$\mathbf{R}_j(k) = \sum_{i=1, i \neq j}^{K} \sqrt{\frac{P_i \rho_{ji}}{M_i N_S}} \mathbf{H}_{ji}(k) \mathbf{X}_i(k) \tag{3}$$

Here $\mathbf{R}_j(k)$ is the co-channel interference from other Tx Nodes, $N_S$ is the number of sub-carriers in the system. $\mathbf{N}_j(k)$ is the additional Gaussian noise with variance $\sigma_N^2$:

$$\sigma_N^2 = \eta_N + 10 \log_{10}(W_S) + F_N \tag{4}$$

$\eta_N$ is the PSD of the thermal noise, $W_S$ is the bandwidth of the subcarrier, and $F_N$ is the noise figure.

In the interference-limited environment of Ad Hoc networks, it is difficult to obtain the exact channel information of co-channel users $\mathbf{H}_{ji}(k), i \neq j$. Instead, we use the MMSE solution at the receiver, where only the covariance information of $\mathbf{R}_j(k)$ is required:

$$E\{\mathbf{R}_j(k)\mathbf{R}_j^H(k)\} = \sum_{i=1, i \neq j}^{K} \frac{P_i \rho_{ji}}{N_S} \mathbf{I}_N \tag{5}$$

That is, the sum of the noise and interference signal $\mathbf{R}_j(k) + \mathbf{N}_j(k)$ can be seen as a white Gaussian noise with zero mean and variance $\sum_{i=1, i \neq j}^{K} \frac{P_i \rho_{ji}}{N_S} + \sigma_N^2$.

The equation in (2) described the received signal without the transceiver impairments. In the following, we will discuss the RF impairment in section III, and the Base-band impairment in section IV.

## III RF Impairments

As we have discussed, in the RF domain, only the phase noise impairment is considered. In this paper, we model the phase noise both at the transmitter and at the receiver. At the transmitter side, assume that the vector to be sent at the $k$th subcarrier of Tx Node $i$ is $\mathbf{X}_i(k)$. Then the additional noise caused by phase noise is denoted as $\mathbf{G}_{TX,i}(k)$, with covariance $\sigma^2_{Tx,i}\mathbf{I}_M$, and $\sigma^2_{Tx,i} = F_{ICI} \cdot P_i / M_i$. The parameter $F_{ICI}$ is determined by the PSD of the phase noise [12]:

$$F_{ICI} \approx \int_{-W_T/2}^{W_T/2} S_\phi(f) df \qquad (6)$$

Here $W_T$ is the bandwidth of all sub-carriers and $W_T = N_S W_S$ ($W_S$ is given in equation (4)). $W_T$ is set as 20MHz in this paper. $S_\phi(f)$ is the PSD of the phase noise, which is adopted from [16]:

$$S_\phi(f) = 10^{-c} + \begin{cases} 10^{-a}, & |f| < f_l \\ 10^{-(f-f_l)\left(\frac{b}{(f_h-f_l)}\right)-a}, & f > f_l \\ 10^{(f+f_l)\left(\frac{b}{(f_h-f_l)}\right)-a}, & f < -f_l \end{cases} \qquad (7)$$

$a = 8.5$, $b = 2$, $c = 12.5$, $f_l = 10kHz$, $f_h = 100kHz$. The resulted $F_{ICI}$ is -31.9 dBc.

At the receiver side, we rewrite equation (2) by adding the additional noise from phase noise:

$$\mathbf{Y}_j(k) = \mathbf{R}_{jj}(k) + \sum_{i=1,i\neq j}^{K} \mathbf{R}_{ji}(k) + \mathbf{G}_{Rx,j}(k) + \mathbf{N}_j(k) \qquad (8)$$

$$\mathbf{R}_{ji}(k) = \mathbf{H}_{ji}(k)\left\{\sqrt{\frac{P_i\rho_{ji}}{M_i N_S}}\mathbf{X}_i(k) + \mathbf{G}_{Tx,i}(k)\right\} \qquad (9)$$

The covariance of Rx side phase noise $\mathbf{G}_{Rx,j}(k)$ relies on the covariance of the received signal $\mathbf{R}_j(k) = \left\{\mathbf{R}_{jj}(k) + \sum_{i=1,i\neq j}^{K} \mathbf{R}_{ji}(k)\right\}$. Since we have $E\{\mathbf{R}_j(k)\mathbf{R}_j^H(k)\} = \sum_{i=1}^{K} \frac{P_i}{N_S}\rho_{ji}(1+F_{ICI})$, the variance of $\mathbf{G}_{Rx,j}(k)$ is given by $\sigma^2_{Rx,j} = \sum_{i=1}^{K} \frac{P_i}{N_S}\rho_{ji}(1+F_{ICI})F_{ICI} \approx \sum_{i=1}^{K} \frac{P_i}{N_S}\rho_{ji}F_{ICI}$.

In order to distinguish the signal and interference components in equation (6), we rewrite the received signal as:

$$\mathbf{Y}_j(k) = \mathbf{H}_{jj}(k) \cdot \sqrt{\frac{P_j \rho_{jj}}{M_j N_S}} \mathbf{X}_j(k) + \mathbf{V}_j(k) \tag{10}$$

$\mathbf{V}_j(k)$ denotes the equivalent white Gaussian noise into the base-band:

$$\mathbf{V}_j(k) = \rho_{jj} \mathbf{H}_{jj}(k) \mathbf{G}_{Tx,j}(k) + \sum_{i=1, i \neq j}^{K} \mathbf{R}_{ji}(k) + \mathbf{G}_{Rx,j}(k) + \mathbf{N}_j(k) \tag{11}$$

The covariance of $\mathbf{V}_j(k)$ is:

$$\begin{aligned}
\sigma_V^2(j) &= \frac{1}{N_S} \left( \sum_{i=1, i \neq j}^{K} P_i \rho_{ji} + P_j \rho_{jj} F_{ICI} + \sum_{i=1}^{K} P_i \rho_{ji} F_{ICI} \right) + \sigma_N^2 \\
&= \frac{1}{N_S} \left( \sum_{i=1, i \neq j}^{K} P_i \rho_{ji} (1 + F_{ICI}) + 2 P_j \rho_{jj} F_{ICI} \right) + \sigma_N^2 \\
&\approx \frac{1}{N_S} \left( \sum_{i=1, i \neq j}^{K} P_i \rho_{ji} + 2 P_j \rho_{jj} F_{ICI} \right) + \sigma_N^2
\end{aligned} \tag{12}$$

As a result, the SINR at the input of the base-band is:

$$SINR_B(j) = \frac{P_j \rho_{jj}}{\sum_{i=1, i \neq j}^{K} P_i \rho_{ji} + \left( 2 P_j \rho_{jj} F_{ICI} + N_S \sigma_N^2 \right)} = \frac{SINR_{in}(j)}{1 + 2 \cdot SINR_{in}(j) \cdot F_{ICI}} \tag{13}$$

$$SINR_{in}(j) = \frac{P_j \rho_{jj}}{\sum_{i=1, i \neq j}^{K} P_i \rho_{ji} + N_S \sigma_N^2} \tag{14}$$

Here $SINR_{in}(j)$ is the SINR at the input of the RF chain. We plot the $SINR_B$ versus the $SINR_{in}$ in Figure 1. The results show that because of the phase noise, the maximum SINR into the base-band is limited at 28.9dB.

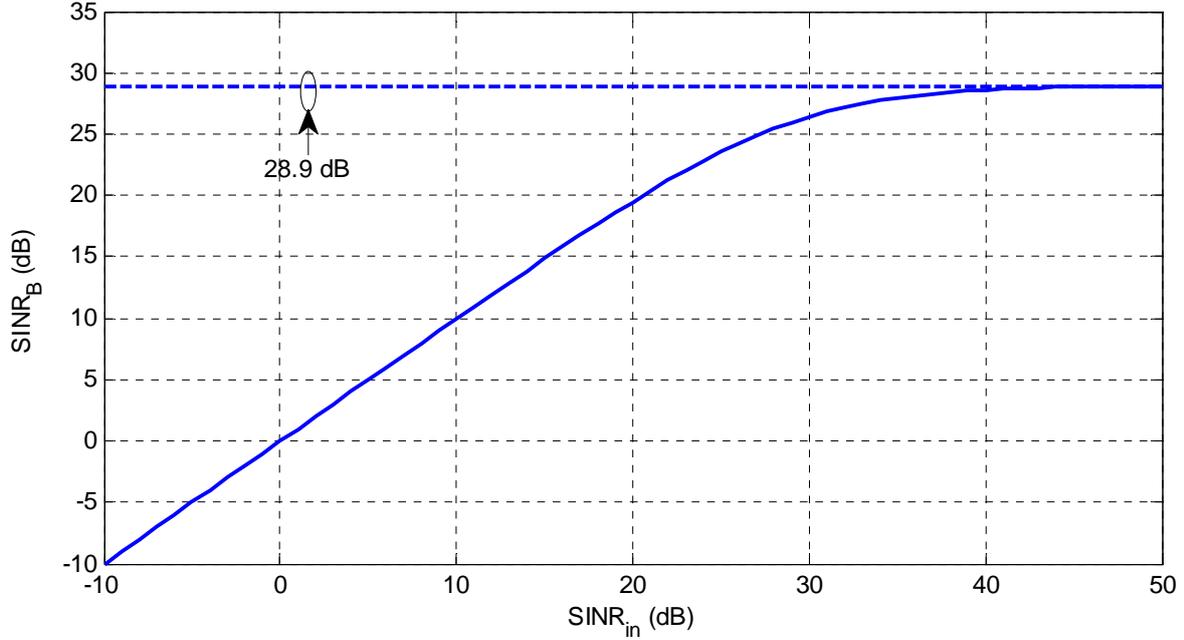

Fig. 1. $SINR_B$ versus the $SINR_{in}$ under phase noise.

## IV. Base-band Impairments

### 1). Residual Frequency Offset

We assume that the normalized frequency offset $\varepsilon$ is within the range of (-0.5, 0.5), and two repeated training symbols are employed to estimate the frequency offset (similar to [13]). The residual frequency offset (RFO) at Rx Node $j$, $\hat{\varepsilon}_j$, is modeled as a Gaussian variable with zero mean and variance $\sigma^2_{RFO,j} = \dfrac{1}{(2\pi)^2 \cdot N_{sub} \cdot SINR_B(j)}$. Consider a specific RFO $\hat{\varepsilon}_j$, as is shown in [13], a lower bound for the degraded SNR is:

$$SINR_{RFO}(\hat{\varepsilon}_j) = \frac{SINR_B(j) \cdot \left\{ \sin \pi\hat{\varepsilon}_j / (\pi\hat{\varepsilon}_j) \right\}^2}{1 + 0.5947 \cdot SINR_B(j) \cdot \left( \sin \pi\hat{\varepsilon}_j \right)^2} \quad (15)$$

We assume that all sub-carriers use the same modulation scheme, which includes BPSK, QPSK, 16QAM and 64QAM. We use $U_j$ to denote the number of bits per modulated symbols at Tx Node $j$, and its value is chosen from 1, 2, 4, and 6. Assume that for an input SNR $SINR_{RFO}(\hat{\varepsilon}_j)$, with $M_j$ transmit antennas and modulation scheme $U_j$, the BER at the MIMO detector is $P_{RFO}\{SINR_{RFO}(\hat{\varepsilon}_j), M_j, U_j\}$, then the average BER over different RFO in the base-band is given by:

$$P_B\{SINR_B(j), M_j, U_j\} = \int_{-\infty}^{\infty} P_{RFO}\{SINR_{RFO}(\hat{\varepsilon}_j), M_j, U_j\} \cdot f(\hat{\varepsilon}_j) d\hat{\varepsilon}_j \qquad (16)$$

$$f(\hat{\varepsilon}_j) = \frac{1}{\sqrt{2\pi\sigma_{RFO,j}^2}} \exp\left\{-\frac{1}{2\sigma_{RFO,j}^2}\hat{\varepsilon}_j^2\right\} \qquad (17)$$

## 2). Imperfect Channel Estimation

In the following, we discuss the evaluation of $P_{RFO}\{SINR_{RFO}(\hat{\varepsilon}_j), M_j, U_j\}$ in equation (12). It involves both the channel estimation and MIMO detection. In terms of the channel estimation, we use an orthogonal training symbols method to estimate the channel between each Tx ant and Rx ant, which is similar to [17]. Here we assume that $M_{Training}$ consecutive training symbols are used for channel estimation, and $M_{Training} = 2^n$, if $2^{n-1} < M \leq 2^n$.

For notational simplicity, given $SINR_{RFO}(\hat{\varepsilon}_j)$, we denote the received signal as:

$$\mathbf{Y}_j(k) = \frac{1}{\sqrt{M_j}} \mathbf{H}_{jj}(k) \mathbf{X}_j(k) + \mathbf{N}_{norm,j}(k) \qquad (18)$$

Here $\mathbf{N}_{norm,j}(k)$ is the normalized white Gaussian noise with variance $\frac{1}{SINR_{RFO}(\hat{\varepsilon}_j)}$.

We use a simple LS estimation for the channel information. That is, $\widehat{\mathbf{H}}_{jj}(k)$ is the equal gain average of all training symbols. As a result, the estimated channel $\widehat{\mathbf{H}}_{jj}(k)$ is:

$$\widehat{\mathbf{H}}_{jj}(k) = \mathbf{H}_{jj}(k) + \sqrt{M_j} \cdot \mathbf{N}_{H,jj}(k) \qquad (19)$$

$\mathbf{N}_{H,jj}(k)$ has a variance of $\frac{1}{M_{Training} \cdot SINR_{RFO}(\hat{\varepsilon}_j)}$.

We assume an MMSE MIMO detector, which has a low complexity in hardware implementation. In this case, the signal is detected using a linear weight vector $\widehat{\mathbf{W}}$

$$\widehat{\mathbf{X}}_j(k) = \widehat{\mathbf{W}}_j(k) \mathbf{Y}_j(k) \qquad (20)$$

$$\widehat{\mathbf{W}}_j(k) = \widehat{\mathbf{H}}_{jj}^H(k)\left(\widehat{\mathbf{H}}_{jj}(k)\widehat{\mathbf{H}}_{jj}^H(k) + \frac{1}{SINR_{RFO}(\hat{\varepsilon}_j)}\mathbf{I}_N\right)^{-1} \quad (21)$$

Now given $\mathbf{H}_{jj}(k)$ and $\widehat{\mathbf{H}}_{jj}(k)$, the expected BER for modulation $S_j$ is:

$$P_{RFO,\mathbf{H},\widehat{\mathbf{H}}}\{SINR_{RFO}(\hat{\varepsilon}_j), M_j, U_j\} = \frac{1}{M_j}\sum_{m=1}^{M_j} B_m\{SINR_{RFO}(\hat{\varepsilon}_j), E_m, U_j\} \quad (22)$$

$$B_m\{SINR_{RFO}(\hat{\varepsilon}_j), E_m, U_j\} = \frac{1}{2^{U_j}}\sum_{l=1}^{2^{U_j}} \left(P_{\text{Re}}(Z_l, U_j) + P_{\text{Im}}(Z_l, U_j)\right) \quad (23)$$

$$E_m = [\mathbf{S}_j(k)]_{m,m} \cdot Z_l \quad (24)$$

$$\sigma^2 = \sum_{l \neq m} \left|[\mathbf{S}_j(k)]_{l,l}\right|^2 + \frac{1}{SINR_{RFO}(\hat{\varepsilon}_j)} \quad (25)$$

$$\mathbf{S}_j(k) = \widehat{\mathbf{W}}_j(k)\mathbf{H}_{jj}(k) \quad (26)$$

Here $Z_l$ is the point in the constellation diagram of modulation $U_j$, which has zero mean and unit variance. $P_{\text{Re}}(Z_l, U_j)$ is calculated according to the number of neighbor points that have minimum distance with $\text{Re}(Z_l)$. If there are two neighbor points that have the minimum distance with $\text{Re}(Z_l)$, then

$$P_{\text{Re}}(Z_l, U_j) = Q\left(\frac{\text{Re}(E_m) - \left(\text{Re}(Z_l) + \frac{1}{\sqrt{D_j}}\right)}{\sqrt{\sigma^2/2}}\right) - Q\left(\frac{\text{Re}(E_m) - \left(\text{Re}(Z_l) - \frac{1}{\sqrt{D_j}}\right)}{\sqrt{\sigma^2/2}}\right) \quad (27)$$

On the other hand, if there is only one neighbor point that has the minimum distance to $\text{Re}(Z_l)$, then

$$P_{\text{Re}}(Z_l, U_j) = Q\left(-\frac{\left|\text{Re}(E_m) - \left(\text{Re}(Z_l) - \text{sgn}(\text{Re}(Z_l))\frac{1}{\sqrt{D_j}}\right)\right|}{\sqrt{\sigma^2/2}}\right) \quad (28)$$

$P_{\text{Im}}(Z_l)$ has the similar form with $P_{\text{Re}}(Z_l)$. $D_j$ is a constant value determined by $U_j$, and $D_j = \{1, 2, 10, 42\}$ corresponding to $U_j = \{1, 2, 4, 6\}$.

The expected BER at the MIMO detector for $SINR_{RFO}(\hat{\varepsilon}_j)$ is written as:

$$P_{RFO}\left\{SINR_{RFO}(\hat{\varepsilon}_j), M_j, U_j\right\} = E_{(\mathbf{H}, \hat{\mathbf{H}})}\left\{P_{RFO,\mathbf{H},\hat{\mathbf{H}}}\left\{SINR_{RFO}(\hat{\varepsilon}_j), M_j, U_j\right\}\right\} \quad (29)$$

Here the expectation is over the random variable $\mathbf{H}$ and $\hat{\mathbf{H}}$.

### 3) Discussions

Now combining the effects of RFO and imperfect channel estimation, the BER $P_B\left\{SINR_B(j), M_j, U_j\right\}$ can be evaluated via equations (16), (22) and (29). In order to verify the efficacy of our derivations, we setup another simulation framework (Fig. 2), which evaluates the performance under RFO and imperfect channel estimation using Monte Carlo simulations. We compare the result from equations ((16), (22) and (29)) and that from Monte Carlo simulations in Fig. 3, where the simulated results approach the analytical ones (equations (16), (22) and (29)) closely. Thus, the efficacy of our derived equations is confirmed.

Furthermore, we evaluate the impacts from residual frequency offset and imperfect channel estimation individually, and the results are shown in Fig. 4. We can see that the RFO's impact is inconsiderable, while the imperfect channel estimation will cause a loss of 2-3dB in SNR. These phenomena are also observed in Monte Carlo simulations. (The results are omitted here for space limitation.)

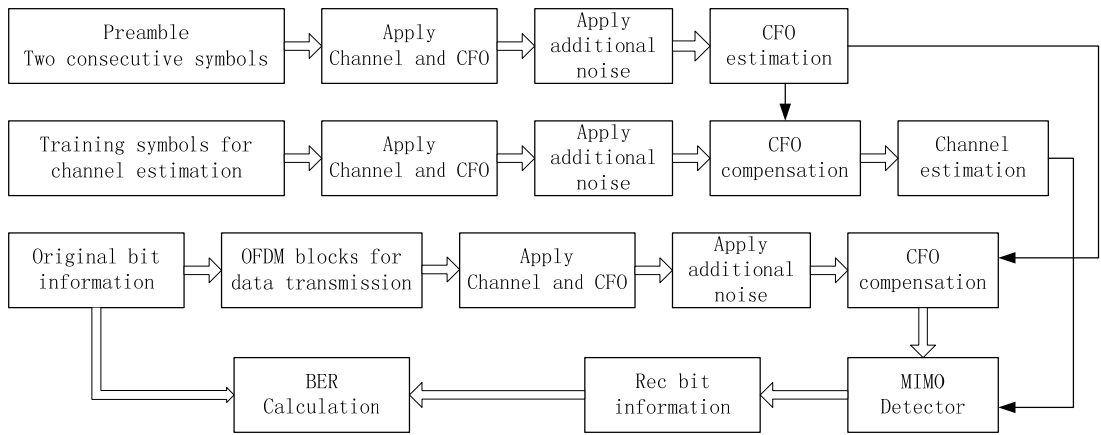

Fig. 2. Monte Carlo simulations for residual frequency offset and imperfect channel estimation.

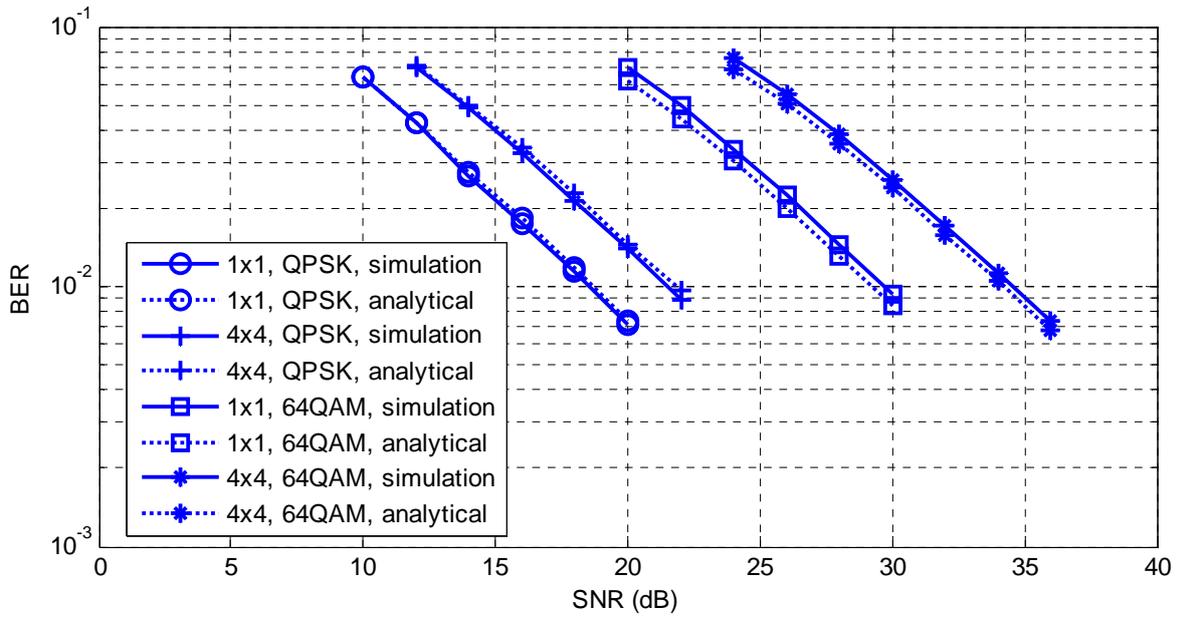

Fig. 3. The comparison between the simulated results and the analytical ones.

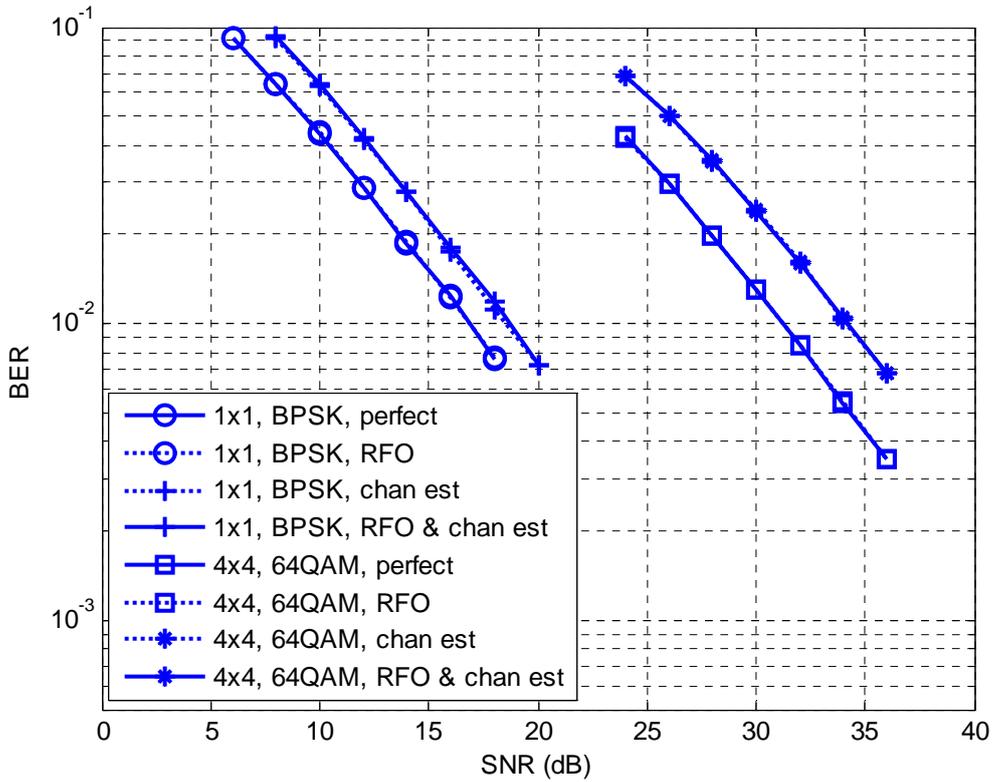

Fig. 4. Effects of the residual frequency offset and imperfect channel estimation.

## V. Throughput Optimization in the Network

### 1). Adaptive Stream Control

Using the results from Section III and IV, now we present the proposed analytical framework in Fig. 5. In this figure, the BER for a specific $SINR_{in}(j)$ under $M_j$ transmit antennas and modulation scheme $U_j$ can be calculated. We define this BER value as $P_{RF}\{SINR_{in}(j), M_j, U_j\}$. Note that $P_{RF}\{SINR_{in}(j), M_j, U_j\} = P_B\{SINR_B(j), M_j, U_j\}$, where $SINR_B(j)$ is derived from $SINR_{in}(j)$ via equation (13).

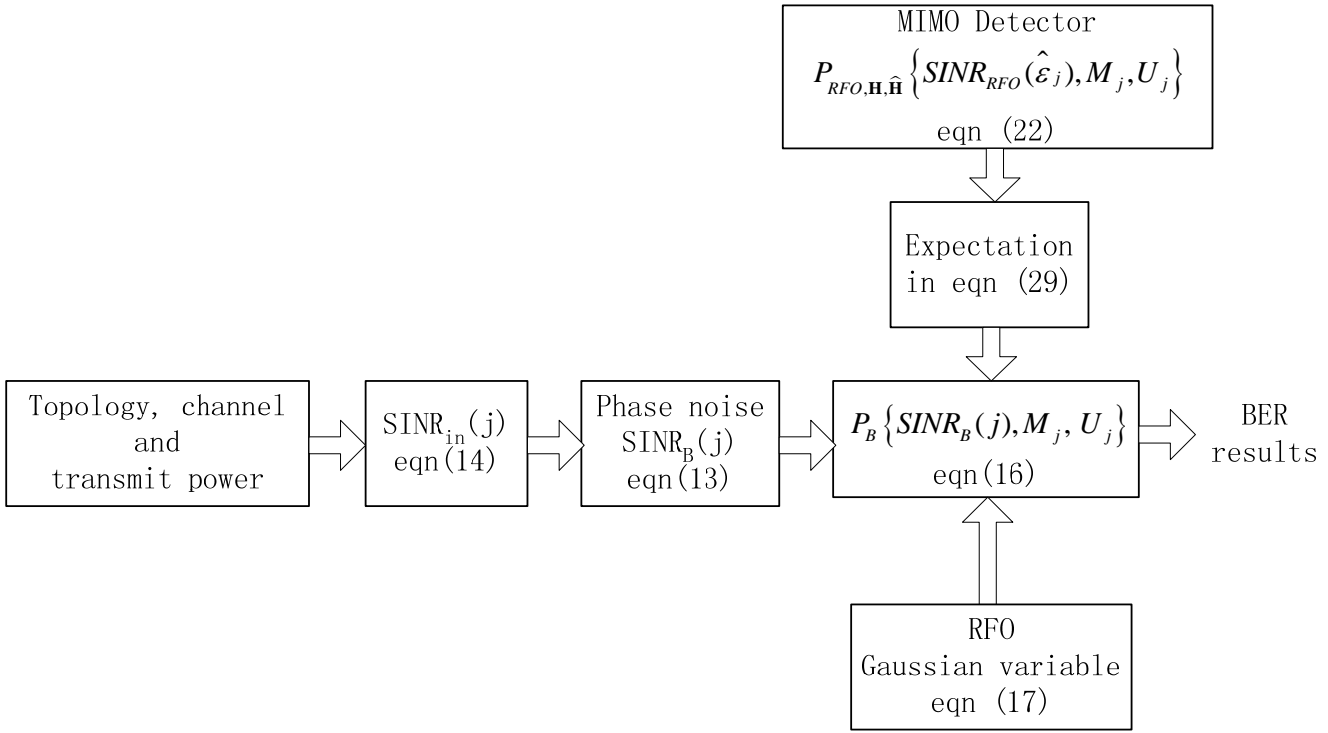

Fig. 5. The proposed analytical framework for transceiver impairments.

In order to evaluate the successful transmission, we define a QoS requirement at the receiver side. That is, the transmitted information is successfully received if and only if the average BER $P_{RF}\{SINR_{in}(j), M_j, U_j\}$ is not smaller than $\Gamma$. We assume a rate 1/2 convolutional code at the base-band. According to Fig. 3 in [18], $\Gamma$ is set as 0.02 in this paper. In this sense, the adaptive stream control is to select the proper parameters $M_j$ and $U_j$ that result in the maximum throughput:

$$\begin{aligned} \max \quad & R \cdot M_j \cdot U_j \\ s.t. \quad & P_{RF}\{SINR_{in}(j), M_j, U_j\} \leq \Gamma \end{aligned} \quad (30)$$

Here $R$ is a constant parameter representing the throughput of a single stream with BPSK modulation. In this paper, since we assume 64 sub-carriers and 20MHz bandwidth, as well as the Rate 1/2 convolution code. We can have $R = 8 Mbps$. The result of equation (30) is denoted as $T_N\{SINR_{in}(j)\}$ in the following, and this result can be obtained via offline method.

Assuming 1 and 4 receive antennas, we plot the $T_N\{SINR_{in}(j)\}$ in Fig. 6. We show that compared with the case that has no impairments, the SINR thresholds under transceiver impairments are increased around 3dB. We note that using 4 receive antennas and without impairments, the SINR threshold for 192Mbps is 30.1dB. However, this mode cannot be supported when the impairments exist in the system. The reason behind is that because of the phase noise, the upper bound of the $SINR_B$ is only 28.9dB (Fig. 1). Thus, we can say that the phase noise will constrain the maximum value of $SINR_B$ and mainly affects the high-rate modes that require high SINR thresholds.

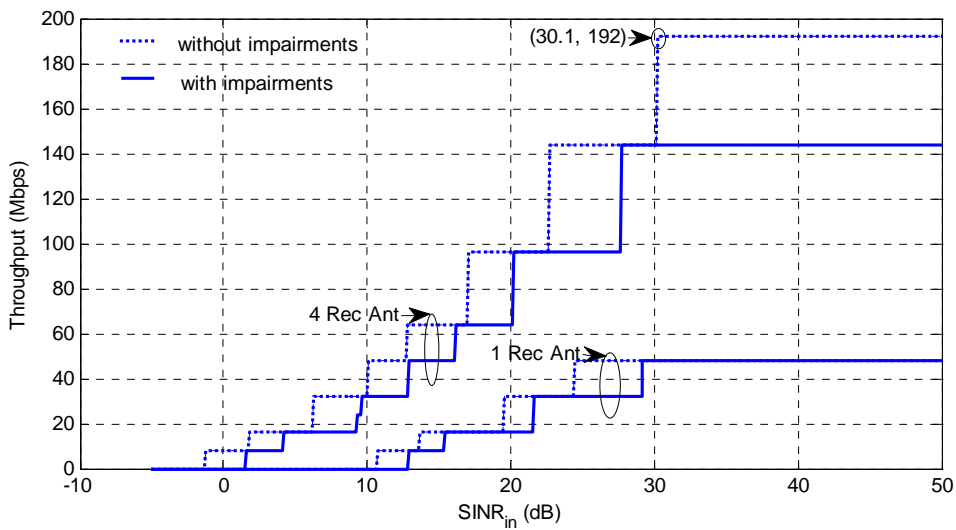

Fig. 6. The results of the adaptive stream control under 1 and 4 receive antennas.

## 2). Maximum Sum Throughput

Assume that Tx Node $j$ uses power $P_j$, and we calculate $SINR_{in}(j)$ using $\{P_j, 1 \le j \le K\}$ via equation (14). Then the maximum sum throughput (MST) in the network is calculated by:

$$\max \sum_{j=1}^{K} T_N\{SINR_{in}(j)\} \quad (31)$$
$$s.t. \quad 0 \le P_j \le P_T$$

This is a nonlinear optimization because $T_N\{\beta\}$ is a nonlinear function. In this paper, we use the Genetic Algorithm (GA) optimization toolbox in MATLAB (2007b) to calculate the equation (31). The size of the population is set as $4 \cdot K$, while other parameters are set as the default values in MATLAB (2007b).

### 3). Distributed Implementation

The MST in equation (31) involves a global optimization that is difficult to be implemented in practical Ad Hoc networks. Here we discuss the distributed control method in Ad Hoc networks that attempts to utilize the enhanced throughput in MIMO OFDM using only local information in each pair. As we have shown in Fig. 5, the optimization in equation (30) will generate a series of data rates with different SINR thresholds. Consequently, the distributed implementation of MST is a distributed power control problem combined with link adaptation process, which is solved conventionally using hard SINR thresholds or soft SINR thresholds. In this paper we jointly use the methods from [19] and [20]. That is, we employ two stages for the power control process. In the first stage, we use the sigmoid function to represent the performance of the system, and obtain a power allocation for each node via iterations. Then in the second stage, we use a hard SNR threshold as well as the adaptive rate selection. The details of the algorithm are given in Appendix I, and the convergence of the provided method is referred to [19], [20] and [21]. Note that the whole algorithm is executed in a distributed manner.

## VI. Simulation Results

We assume that Tx Nodes are identically distributed in a disk with 1000m. In each transceiver pair, the distance between the Tx Node and the Rx Node is uniformly distributed between 10m and 300m. Simulation results are averaged from 1000 trials. Other parameters used in the simulations are collected in Table I.

Table I. Parameters in the simulation.

| Parameter | Value |
|---|---|
| Path loss | $d_0 = 1m$ <br> $L_P(d_0) = -46dB$ <br> $\alpha = 3$ |
| Noise PSD  $\eta_N$ | -174dBm/Hz |
| Bandwidth of sub-carrier  $W_S$ | 312.5kHz |
| Number of sub-carriers  $N_S$ | 64 |
| Noise Figure  $F_N$ | 4dB |
| Bandwidth of all sub-carriers  $W_T$ | 20MHz |
| Phase Noise  $F_{ICI}$ | -31.9dBc |
| SINR threshold  $\Gamma$ | 0.02 |
| Maximum Transmit Power  $P_T$ | 20dBm |

We first evaluate the MST under 1 to 4 receive antennas. We present both the results without transceiver impairments (solid lines), and those with impairments (dotted lines) in Fig. 7. The results clearly show that the MST is greatly improved by using multiple antennas. Meanwhile, this improvement is inevitable degraded by the transceiver impairments and such loss is shown as non-negligible in Fig. 7. We also note that the MST increases with more transceiver pairs. This is a natural result of the multi-user diversity. Moreover, we extend the simulations with up to 8 antennas and plot the results in Fig. 8. As expected, the MST improvement with multiple antennas and the loss caused by transceiver impairments are demonstrated again, especially when the number of receive antennas is large.

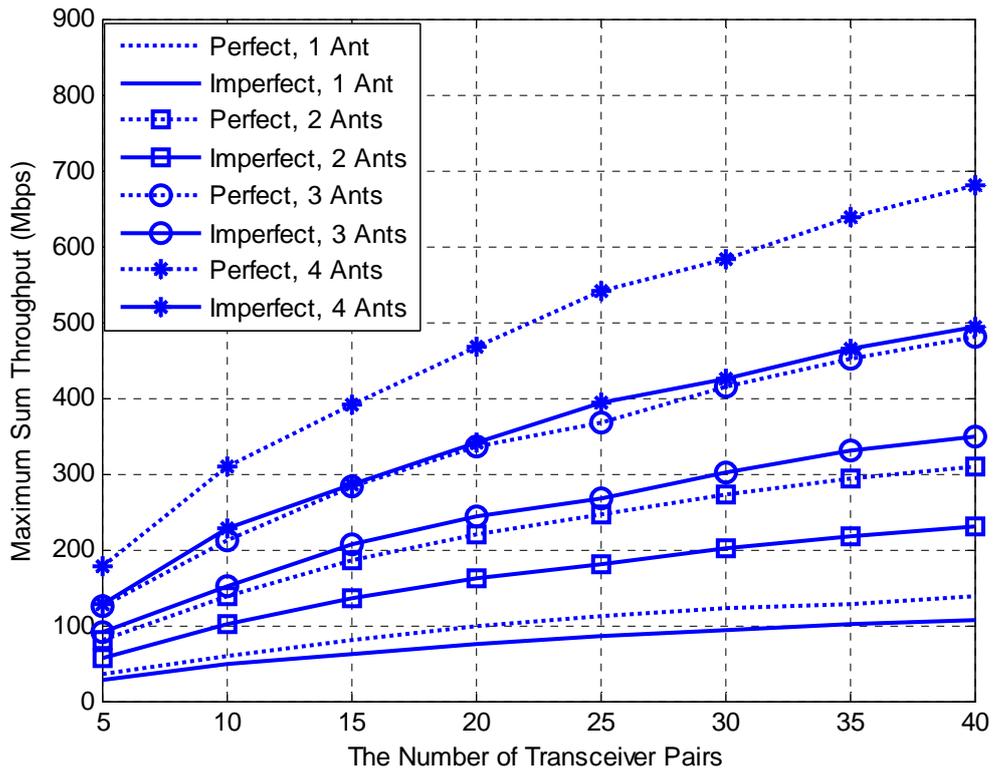

Fig. 7 The MST under 1 to 4 receive antennas.

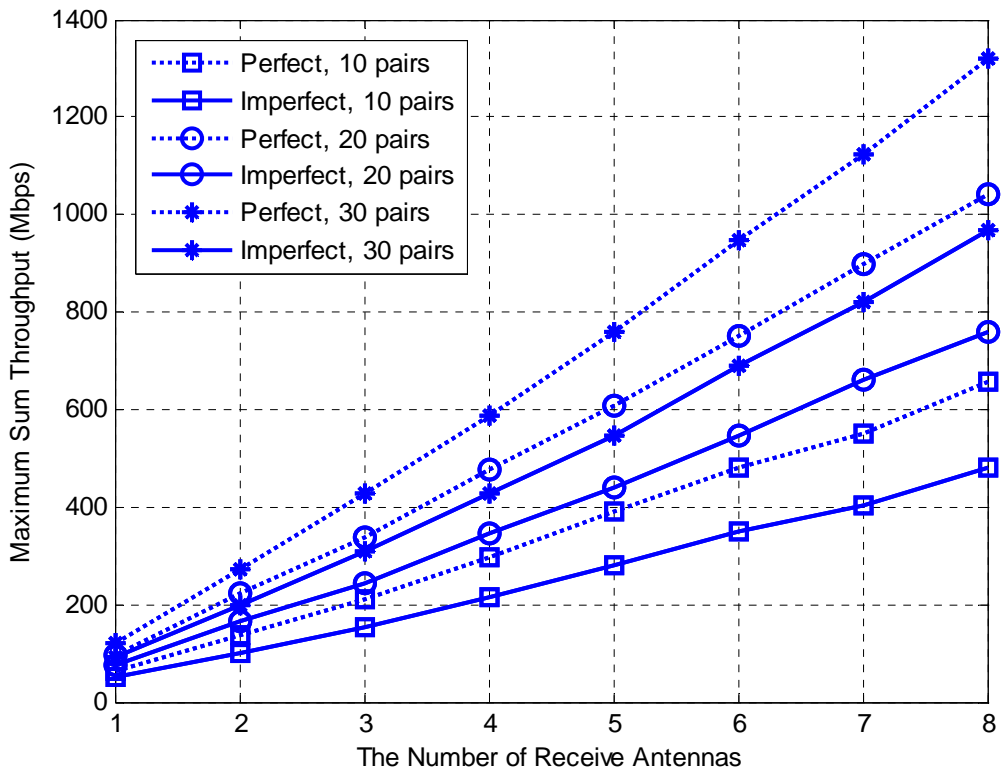

Fig. 8 MST versus different number of receive antennas.

Next, in Fig. 9, we numerically scale the loss in MST due to transceiver impairments. 4 receive antennas at each pair is used. Assume that the MST with transceiver impairments is $T_w$, while that without impairments is $T_{wo}$. Then we define the loss ratio in MST as $\frac{T_{wo}-T_w}{T_{wo}}$. We show that for the simulated parameters, the loss ratio from all the considered impairments is around 27%. We also evaluate the loss ratio from different impairments individually. The results indicate that RFO only introduces a negligible impact. For the phase noise, we have demonstrated in Section II that the phase noise only constrains the maximum SINR into the base-band, thus it mainly affects the high-rate mode requiring high SINR threshold. In our case, the loss ratio from phase noise is about 7%. Finally, imperfect channel estimation has the largest loss ratio, which is 22% in Fig. 9.

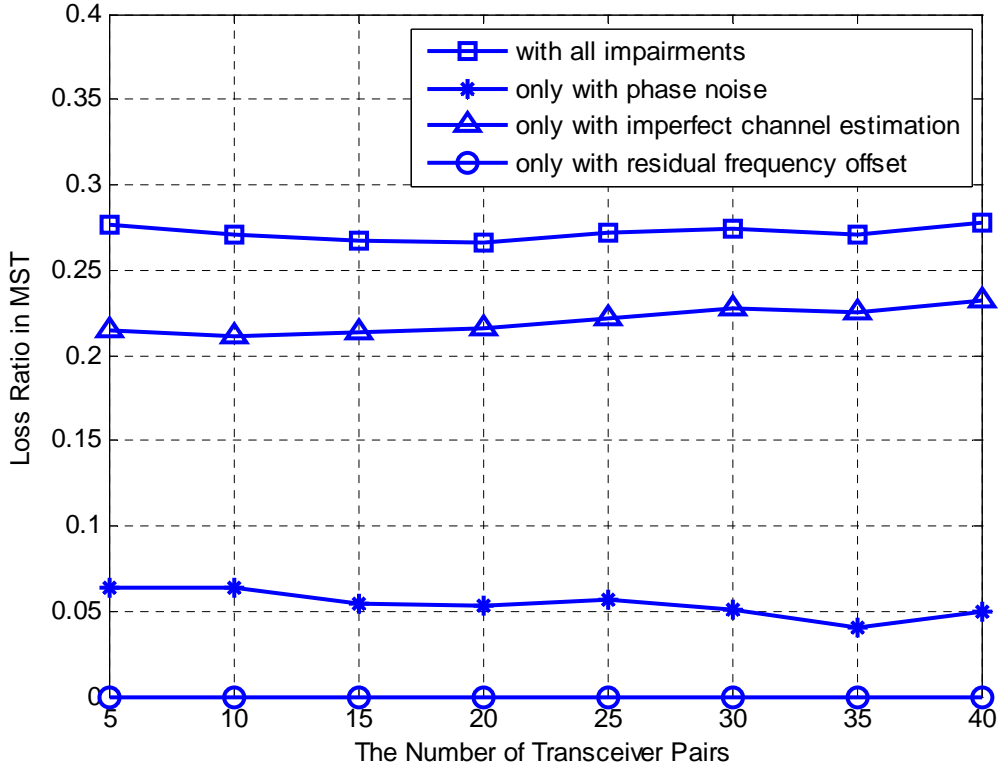

Fig. 9 Loss ratio for different impairments.

The sum throughput using the Distributed Power and Rate Control (DPRC) method is shown in Fig. 10, where the MST results are also provided. Since only local information is used in DPRC, the sum throughput is definitely lower than the MST. However, in this distributed method, we can also observe the enhancement in the sum throughput, which is from multiple antennas. Meanwhile, similar to the MST case in Fig. 7, the transceiver impairments also significantly affect the results of the DPRC. Finally, we point out that the sum throughput with DPRC increases with more transceiver pairs, thus multi-user diversity is explored here.

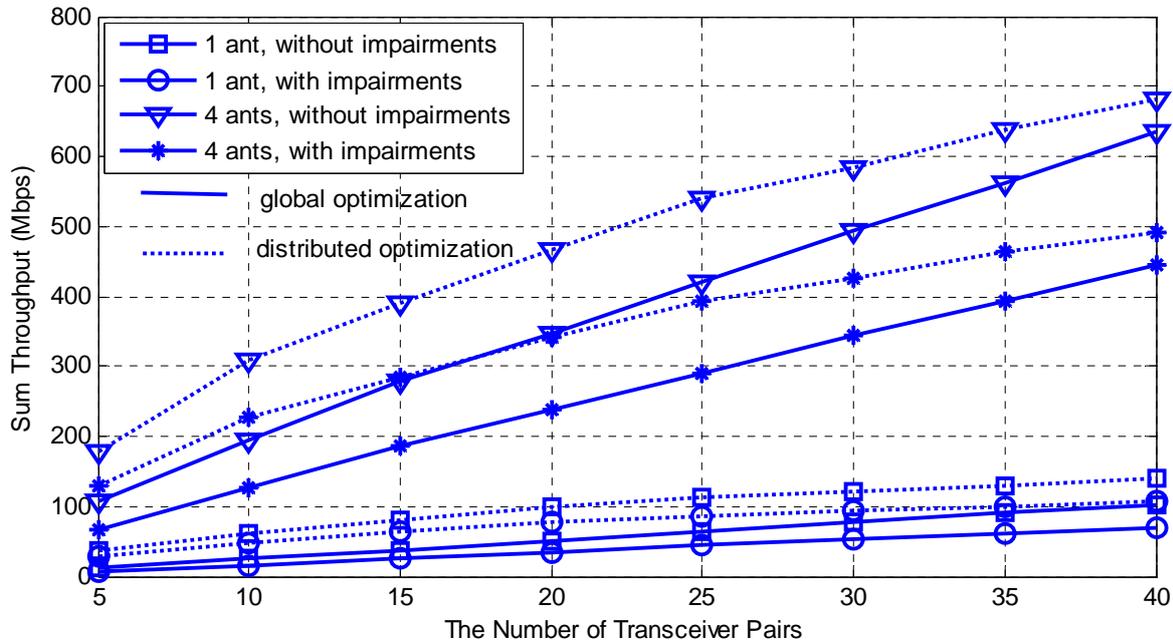

Fig. 10 Sum throughput using distributed power and rate control method.

## VII. Conclusion

This paper presented an analytical framework to evaluate the impacts of transceiver impairments in MIMO-OFDM based Ad Hoc networks. The considered impairments include phase noise, residual frequency offset and imperfect channel estimation. Numerical equations are provided to scale the effect of impairments, as well as the performance of the whole system.

We firstly evaluate the Maximum Sum Throughput (MST) in the network. For the simulated parameters, when 4 receive antennas are used, the results show a loss ratio of 30% in the MST that is due to the transceiver impairments. Among these impairments, RFO only introduces a negligible impact, while imperfect channel estimation seriously degrades the MST. Phase noise constrains the maximum SINR into the base-band and mainly affects the high-rate modes that require high SINR thresholds. We further evaluate the impacts of transceiver impairments in the distributed power and rate control method that aims to maximize the sum throughput using only local information, and non-negligible loss due to the impairments still exists.

Although only the sum throughput is considered in this paper, our proposed framework can be easily extended to other performance metrics in Ad Hoc network. For instance, the fair throughput in the network can be evaluated using our analytical framework and the methods in [22] (global optimization) and [23] (distributed implementation). This is also our future research topic.

# Reference

[1] Biao Chen; Gans, M.J., "MIMO communications in ad hoc networks," Signal Processing, IEEE Transactions on [see also Acoustics, Speech, and Signal Processing, IEEE Transactions on] , vol.54, no.7, pp. 2773-2783, July 2006

[2] Ioannis Sarris; Andrew R. Nix; Angela Doufexi, "High-throughput multiple-input multipleoutput systems for in-home multimedia streaming," Wireless Communications, IEEE [see also IEEE Personal Communications] , vol.13, no.5, pp.60-66, October 2006

[3] Winters, J.H., "Smart antenna techniques and their application to wireless ad hoc networks," Wireless Communications, IEEE [see also IEEE Personal Communications] , vol.13, no.4, pp. 77-83, Aug. 2006

[4] Sigen Ye; Blum, R.S., "On the rate regions for wireless MIMO ad hoc networks," Vehicular Technology Conference, 2004. VTC2004-Fall. 2004 IEEE 60th , vol.3, no., pp. 1648-1652 Vol. 3, 26-29 Sept. 2004

[5] Liang, C.; Dandekar, K.R., "Power Management in MIMO Ad Hoc Networks: A Game-Theoretic Approach," Wireless Communications, IEEE Transactions on , vol.6, no.4, pp.1164-1170, April 2007

[6] Baccarelli, E.; Biagi, M.; Pelizzoni, C.; Cordeschi, N., "Optimized Power Allocation for Multiantenna Systems Impaired by Multiple Access Interference and Imperfect Channel Estimation," Vehicular Technology, IEEE Transactions on , vol.56, no.5, pp.
3089-3105, Sept. 2007

[7] Govindasamy, S.; Bliss, D.W.; Staelin, D.H., "Spectral Efficiency in Single-Hop Ad-Hoc Wireless Networks with Interference Using Adaptive Antenna Arrays," Selected Areas in Communications, IEEE Journal on , vol.25, no.7, pp.1358-1369, September 2007

[8] Ling, J.; Chizhik, D.; Samardzija, D.; Valenzuela, R.A., "Peer-to-Peer MIMO Radio Channel Measurements in a Rural Area," Wireless Communications, IEEE Transactions on , vol.6, no.9, pp.3229-3237, September 2007

[9] Living with a real radio: Impact of front-end effects. The International Series in Engineering and Computer Science, ISSN: 0893-3405. Volume 692: Wireless OFDM Systems. Springer Netherlands.

[10] Hyun Woo Kang; Yong Soo Cho; Dae Hee Youn, "On compensating nonlinear distortions of an OFDM system using an efficient adaptive predistorter," Communications, IEEE Transactions on , vol.47, no.4, pp.522-526, Apr 1999.

[11] Tarighat, A.; Sayed, A.H., "MIMO OFDM Receivers for Systems With IQ Imbalances," Signal Processing, IEEE Transactions on , vol.53, no.9, pp. 3583-3596, Sept. 2005.

[12] Liang-Hui Li; Fu-Lin Lin; Huey-Ru Chuang, "Complete RF-System Analysis of Direct Conversion Receiver (DCR) for 802.11a WLAN OFDM System," Vehicular Technology, IEEE Transactions on , vol.56, no.4, pp.1696-1703, July 2007.

[13] Moose, P.H., "A technique for orthogonal frequency division multiplexing frequency offset correction," Communications, IEEE Transactions on , vol.42, no.10, pp.2908-2914, Oct 1994.

[14] Bin Xia; Jiangzhou Wang; Sawahashi, M., "Performance comparison of optimum and MMSE receivers with imperfect channel estimation for VSF-OFCDM systems," Wireless Communications, IEEE Transactions on , vol.4, no.6, pp. 3051-3062, Nov. 2005.

[15] T. S. Rappaport, Ed., Wireless Communications: Principles & Practice. Englewood Cliffs, NJ: Prentice-Hall, 1996.

[16] Guanghui Liu; Weile Zhu, "Compensation of phase noise in OFDM systems using an ICI reduction scheme," Broadcasting, IEEE Transactions on , vol.50, no.4, pp. 399-407, Dec. 2004.

[17] IEEE P802.11n/D3.00, Wireless LAN Medium Access Control (MAC) and Physical Layer (PHY) specifications: Amendment 4: Enhancements for Higher Throughput.

[18] Fei Peng; Jinyun Zhang; Ryan, W.E., "Adaptive Modulation and Coding for IEEE 802.11n," Wireless Communications and Networking Conference, 2007.WCNC 2007. IEEE , vol., no., pp.656-661, 11-15 March 2007

[19] Huang, W. L.; Letaief, K. B., "Cross-Layer Scheduling and Power Control Combined With Adaptive Modulation for Wireless Ad Hoc Networks," Communications, IEEE Transactions on , vol.55, no.4, pp.728-739, April 2007

[20] Ginde, S.S. V.; Mackenzie, A.A. B.; Buehrer, R.R. M.; Komali, R.R. S., "A Game-Theoretic Analysis of Link Adaptation in Cellular Radio Networks", Vehicular Technology, IEEE Transactions on : Accepted for future publication.

[21] Mingbo Xiao; Shroff, N.B.; Chong, E.K.P., "A utility-based power-control scheme in wireless cellular systems," Networking, IEEE/ACM Transactions on , vol.11, no.2, pp. 210-221, Apr 2003

[22] Chiang, M.; Chee Wei Tan; Palomar, D.P.; O'Neill, D.; Julian, D., "Power Control By Geometric Programming," *Wireless Communications, IEEE Transactions on* , vol.6, no.7, pp.2640-2651, July 2007

[23] Wang, K.; Proakis, J.G.; Rao, R.R., "Distributed fair scheduling and power control in wireless ad hoc networks," *Global Telecommunications Conference, 2004. GLOBECOM '04. IEEE* , vol.6, no., pp. 3557-3562, 29 Nov.-3 Dec. 2004

## Appendix I    Distributed Power and Rate Control (DPRC) Algorithm

In the description of the algorithm, we assume that the power for the $i$ th transceiver power is $P_i$. The parameter $Loop\_Num$ is set as 30. The provided distributed power and rate control method is composed of two stages.

1. In the first stage, we use the sigmoid function to model the performance of the system [21]:

$$S_i(SINR) = \frac{1}{1+e^{-a(SINR-\beta)}} \quad (32)$$

The objective function to be optimized is:

$$T_i(SINR) = \frac{1}{1+e^{-a(SINR-\beta)}} - \alpha P_i \quad (33)$$

$a = 1$, $\alpha = 0.001$, $\Gamma = 1.001$, $\beta = \Gamma - \frac{1}{a}\ln(a\Gamma - 1)$.

The power control is executed using the game theory that is similar to those in [19] and [21]:

1.1 $P_i(0) = \alpha P_T$. $\alpha$ is uniformly distributed in [0, 1].

1.2 $loop = 1$

1.3 For each user $i$

1.3.1 $I_i(loop) = \sum_{l=1,l \neq i}^{K} P_l(loop)\rho_{il} + \sigma_N^2$. If $\frac{\alpha I_i}{\rho_{ii}} < \frac{a}{4}$, then

$$P_i(loop) = \beta \frac{I_i(loop)}{\rho_{ii}} - \frac{I_i(loop)}{a\rho_{ii}} \ln\left(\frac{a}{2\alpha \frac{I_i(loop)}{\rho_{ii}}} - 1 - \sqrt{\left(\frac{a}{2\alpha \frac{I_i(loop)}{\rho_{ii}}} - 1\right)^2 - 1}\right); \text{ otherwise, } P_i(loop) = 0.$$

1.3.2 If $P_i(loop) > P_T$, $P_i(loop) = P_T$.

1.3.3 $SINR_T(i,loop) = P_i(loop)\frac{\rho_{ii}}{I_i(loop)}$. If

$$\frac{1}{1+\exp(-a \cdot (SINR_T(i,loop) - \beta))} - \alpha P_i(loop) < \frac{1}{1+\exp(a\beta)}, \text{ then } P_i(loop) = 0.$$

1.4 If $loop < Loop\_Num$, then $loop = loop + 1$ and go to step 1.3; else, go to step 2.

2. In the second stage, we re-adjust the power and select the proper rate at the same time. We define these SINR thresholds as $\{\gamma_q, 1 \leq q \leq Q\}$, corresponding $Q$ different data rates in ascending order.

2.1 $P_i(0) = P_i(Loop\_Num)$, $loop = 1$.

2.2 for each user $i$

2.2.1 $SINR_T(i, loop) = P_i(loop) \frac{\rho_{ii}}{I_i(loop)}$, if $SINR_T(loop) < \gamma_1$, $R(i) = 0$; else, find the maximum $q$ that satisfies $SINR_T(i) \geq \gamma_q$, set $R(i) = q$.

2.3 for each user $i$

2.3.1 If $R(i) > 0$, $P_i(loop+1) = P_i(loop) \frac{SINR_T(i, loop)}{\gamma_{R(i)}}$.

2.4 If $loop < Loop\_Num$, then $loop = loop + 1$ and go to step 2.2; else, go to step 3.

3. Now transceiver pair $i$ will use the power $P_i(Loop\_Num)$, and select the highest rate that satisfies equation (30).